\documentclass[runningheads]{llncs}
\usepackage{graphicx}
\usepackage{amsmath}
\usepackage{amssymb}

\DeclareSymbolFont{matha}{OML}{txmi}{m}{it}
\DeclareMathSymbol{\varv}{\mathord}{matha}{118}
\usepackage{array}
\usepackage{tabularx}
\usepackage{multirow, booktabs}
\usepackage{float}
\usepackage{bm}
\usepackage[breaklinks=true,bookmarks=false]{hyperref}
\newcommand\Tstrut{\rule{0pt}{2.5ex}}
\newcolumntype{C}[1]{>{\centering\arraybackslash}p{#1}}

\begin{document}
%
\title{Robust Image Registration with Absent Correspondences in Pre-operative and Follow-up Brain MRI Scans of Diffuse Glioma Patients}
%
\titlerunning{Unsupervised Deformable Image Registration with Absent Correspondences}
%
%
%

\author{Tony C. W. Mok \and Albert C. S. Chung}


\institute{The Hong Kong University of Science and Technology, Hong Kong, China
\email{\{cwmokab,achung\}@cse.ust.hk}}


\maketitle              
\begin{abstract}
Registration of pre-operative and follow-up brain MRI scans is challenging due to the large variation of tissue appearance and missing correspondences in tumour recurrence regions caused by tumour mass effect. Although recent deep learning-based deformable registration methods have achieved remarkable success in various medical applications, most of them are not capable of registering images with pathologies. In this paper, we propose a 3-step registration pipeline for pre-operative and follow-up brain MRI scans that consists of 1) a multi-level affine registration, 2) a conditional deep Laplacian pyramid image registration network (cLapIRN) with forward-backward consistency constraint, and 3) a non-linear instance optimization method. We apply the method to the Brain Tumor Sequence Registration (BraTS-Reg) Challenge. Our method achieves accurate and robust registration of brain MRI scans with pathologies, which achieves a median absolute error of 1.64 mm and 88\% of successful registration rate in the validation set of BraTS-Reg challenge. Our method ranks 1$^\text{st}$ place in the 2022 MICCAI BraTS-Reg challenge.

\end{abstract}

\section{Introduction}
Registration of pre-operative and follow-up images is crucial in evaluating the effectiveness of treatment for patients suffering from diffuse glioma. However, this registration problem is challenging due to the missing correspondences and mass effect caused by resected tissue. While many recent deep learning-based deformable registration algorithms \cite{de2017end,balakrishnan2018unsupervised,dalca2018unsupervised,kim2019unsupervised,hu2019dual,mok2020fast,mok2020large2,mok2021conditional2} are available, only a few learning-based methods \cite{han2020deep} address the missing correspondences problem. In this paper, we propose a 3-step registration pipeline for pre-operative and follow-up brain MRI scans that consists of 1) a multi-level affine pre-alignment, 2) a conditional deep Laplacian pyramid image registration network (cLapIRN) with forward-backward consistency constraint \cite{mok2022unsupervised,mok2021conditional,mok2020large}, and 3) a non-linear instance optimization with inverse consistency. We validate the method using the pre-operative and follow-up images brain MRI scans in the Brain Tumor Sequence Registration Challenge (BraTS-Reg) challenge \cite{baheti2021brain}. 



\section{Related Work}
Accurate registration of pre-operative and post-recurrence brain MRI scans is crucial to the treatment plan and diagnosis of intracranial tumors, especially brain gliomas \cite{heiss2011multimodality,price2007predicting}. To better interpret the location and extent of the tumor and its biological activity after resection, the dense correspondences between pre-operative and follow-up structural brain MRI scans of the patient first need to be established. However, deformable registration between the pre-operative and follow-up scans, including post-resection and post-recurrence, is challenging due to possible large deformations and absent correspondences caused by tumor's mass effects \cite{dean1990gliomas}, resection cavities, tumor recurrence and tissue relaxation in the follow-up scans.  While recent deep learning-based deformable registration (DLDR) methods \cite{balakrishnan2018unsupervised,kim2019unsupervised,hu2019dual,heinrich2019closing,mok2020fast} have achieved remarkable registration performance in many medical applications, these registration approaches often ignored the absent correspondence problem in the pre-operative and post-recurrence images. To address this issue, we extend our deep learning-based method described in \cite{mok2022unsupervised} by introducing affine pre-alignment and non-linear instance optimization as post-processing to our method. DIRAC leverages conditional Laplacian Pyramid Image Registration Networks (cLapIRN) \cite{mok2021conditional} as the backbone network, jointly estimates the bidirectional deformation fields and explicitly locates regions with absent correspondence. By excluding the regions with absent correspondence in the similarity measure during training, DIRAC improves the target registration error of landmarks in pre-operative and follow-up images, especially for those near the tumour regions.

\section{Methods}
We propose a 3-step registration pipeline for pre-operative and follow-up brain MRI scans which consists of 1) a gradient descent-based affine registration method, 2) a deformable image registration method with absent correspondence (DIRAC), and 3) a non-linear instance optimization method. Let $B$ and $F$ be the pre-operative (baseline) scan $B$ and post-recurrence (follow-up) scan defined over a $n$-D mutual spatial domain $\Omega \subseteq \mathbb{R}^n$. Our goal is to establish a dense non-linear correspondence between the pre-operative scan and the post-recurrence scan of the same subject. In this paper, we focus on 3D registration, i.e., $n = 3$ and $\Omega \subseteq \mathbb{R}^3$.

\begin{figure}[t]
	\centering
	\includegraphics[width=1.0\linewidth]{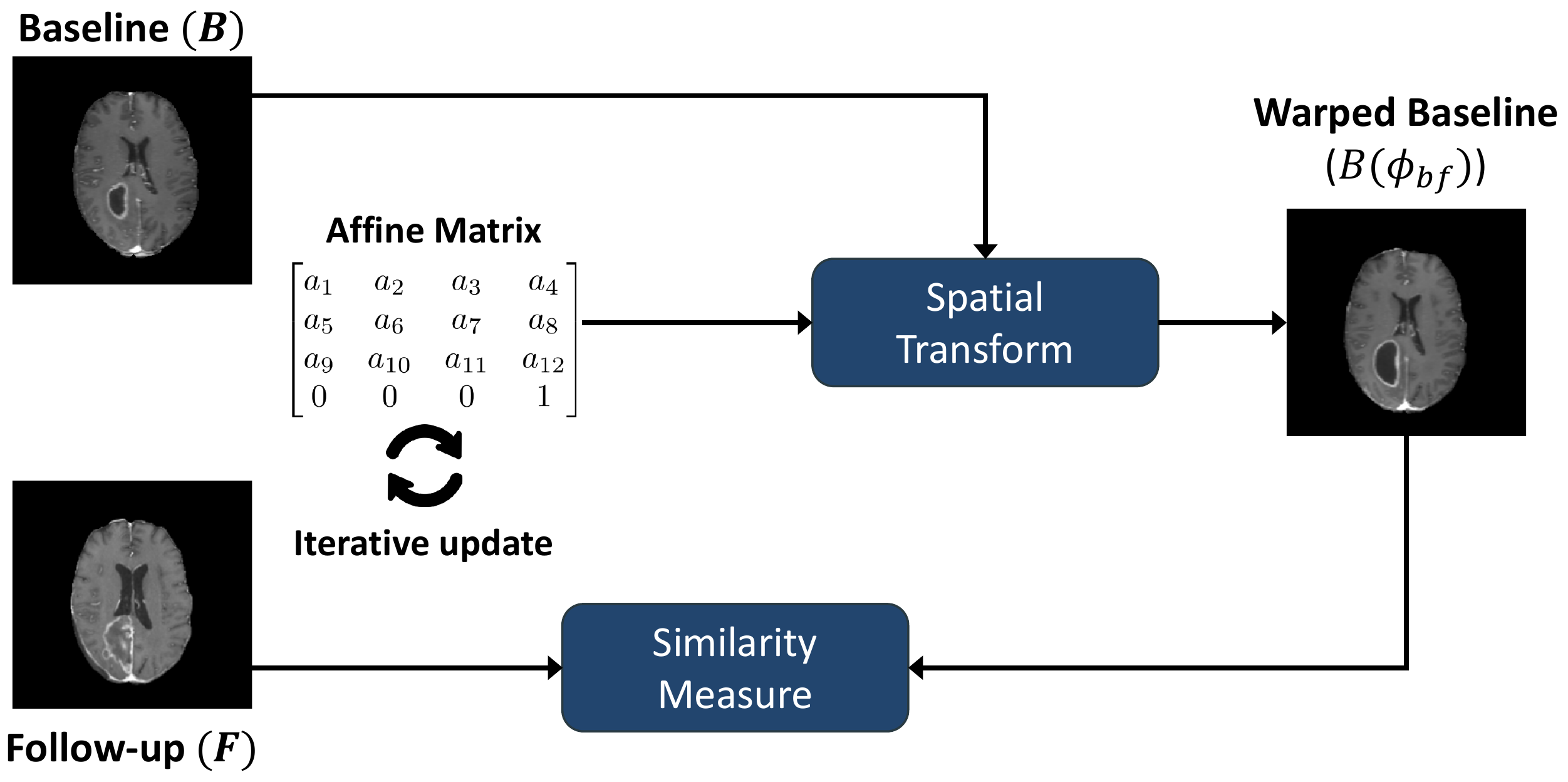}
	\caption{Overview of the affine registration (Step 1). Our method optimizes the affine matrix using instance optimization. Only the baseline scan registered to the follow-up scan is shown for brevity.} \label{fig:affine}
\end{figure}

\subsection{Affine Registration}
Although all MRI scans provided by the challenge are rigidly registered to the same anatomical template \cite{baheti2021brain}, we found that there are large linear misalignments between the pre-operative and follow-up images in cases suffering from serious tumor mass effect. To factor out the possible linear misalignment between MRI scans $B$ and $F$, we register T1-weighted $B$ and $F$ scans using the iterative affine registration method.

Figure \ref{fig:affine} depicts the overview of the affine registration. The affine registration method starts by initializing two identity matrices as initial affine transformation and creating image pyramids with $N_{level}$ levels using trilinear interpolation for $B$ and $F$. Then, we iteratively optimize the solutions by minimizing a suitable distance measure that quantifies alignment quality using the Adam optimizer \cite{kingma2014adam} and a coarse-to-fine multi-resolution iteratively registration scheme. In this step, we use the Normalized Gradient Fields (NGF) distance measure \cite{hodneland2013normalized}. Formally, the NGF is defined as:

\begin{equation}\label{eq:ngf}
	\text{NGF}(B, F) = \int_{\Omega} 1 - \frac{\langle \nabla B, \nabla F \rangle^2}{||\nabla B||_{\epsilon}^2 ||\nabla F||_{\epsilon}^2}
\end{equation}

\noindent where $\langle x, y \rangle := x^\top y$, $||x||_{\epsilon} = \sqrt{x^\top x + \epsilon}$ and $\epsilon$ is an edge parameter controlling the level of influence of image gradients. The value of NGF is minimized when the gradients of $B$ and $F$ are aligned. The result with the minimal distance measure is selected as intermediate result.

\begin{figure}[t]
	\centering
	\includegraphics[width=1.0\linewidth]{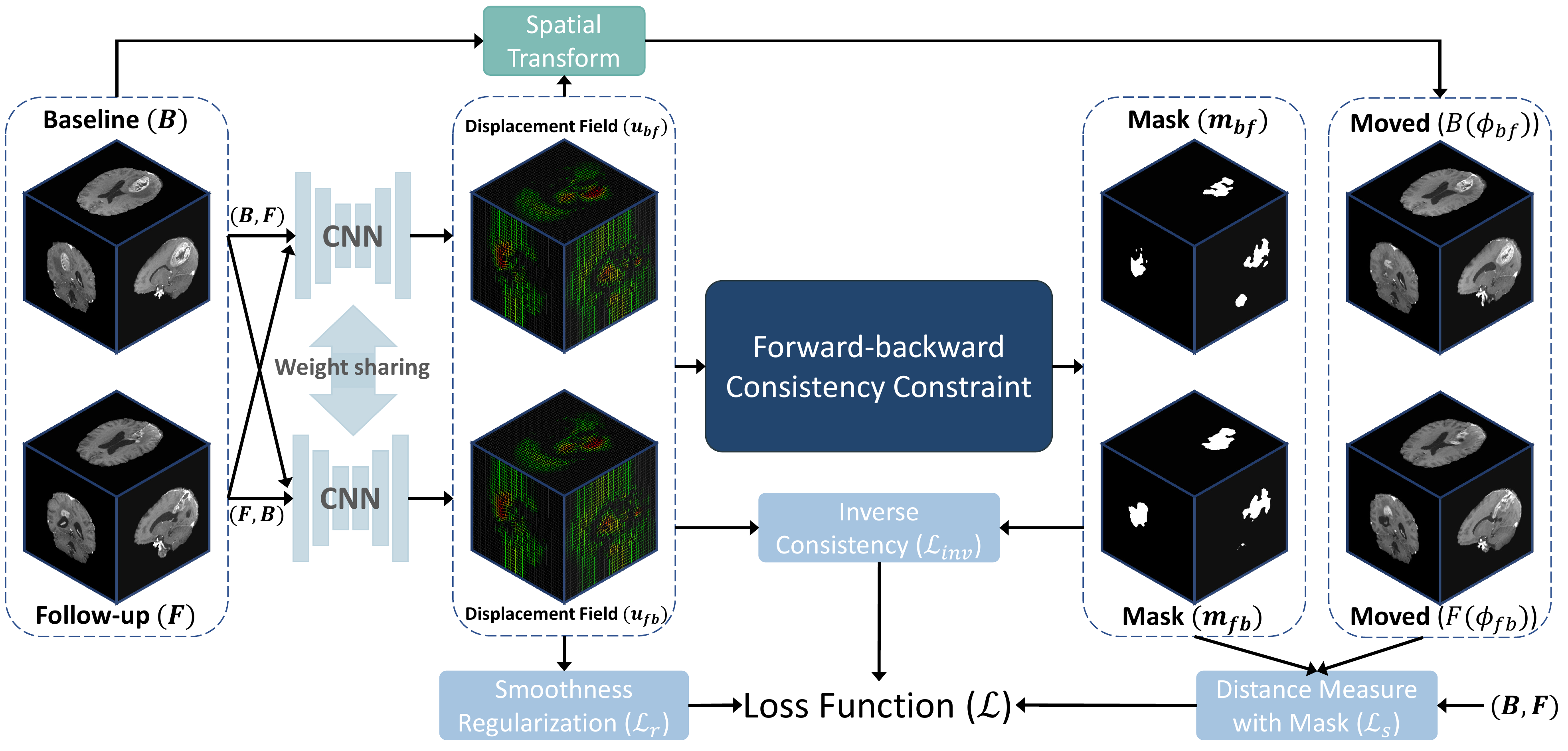}
	\caption{Overview of the proposed deformable image registration method with absent correspondence. Our method jointly estimates the bidirectional deformation fields and locates regions with absent correspondence (denoted as mask). The regions with absent correspondence are excluded in the similarity measure during training. For brevity, the magnitude loss of the masks is omitted in the figure.} \label{fig:unsupervised}
\end{figure}

\subsection{Unsupervised Deformable Registration with Absent Correspondences Estimation}
Assume that baseline scan $B$ and follow-up scan $F$ are affinely aligned in step 1 such that the main source of misalignment between $B$ and $F$ is non-linear, we then apply DIRAC \cite{mok2022unsupervised} to further align $B$ and $F$ in an bidirectional manner. Since multi-parametric MRI sequences of each time-point, including T1 contrast-enhanced (T1ce), T2, T2 Fluid Attenuated (Flair) and T1-weighted (T1), are provided for each case in the BraTS-Reg challenge, we utilize all MRI modalities of the brain MRI scans in this step, i.e., $B$ and $F$ are 4-channel pre-operative and post-recurrence scans, respectively. Specifically, we parameterize the problem with a function $f_\theta(F, B) = (\bm{u}_{fb}, \bm{u}_{bf})$ cLapIRN \cite{mok2021conditional}, where $\theta$ is a set of learning parameters and $\bm{u}_{bf}$ represents the displacement field that transform $B$ to align with $F$, i.e., $B(x+\bm{u}_{bf}(x))$ and $F(x)$ define similar anatomical locations for each voxel $x\in\Omega$ (except voxels with absent correspondence). Figure \ref{fig:unsupervised} illustrates the overview of DIRAC. DIRAC leverages the bidirectional displacement fields and a forward-backward consistency locate regions with absent correspondence and excludes them in the similarity measure during training phase.

\begin{table}[t]
\centering	
\caption{Parameters used in the registration pipeline. NCC-SP: Negative Local Cross-correlation with Similarity Pyramid.}\label{tab:parameters}
\resizebox{\textwidth}{!}{%
\begin{tabular}{c|c|c|c}
\multirow{2}{*}{\textbf{Parameters}}  & \multicolumn{3}{c}{\textbf{Methods}} \\ \cline{2-4} 
                             & \textbf{Affine}\Tstrut & \textbf{DIRAC} & \textbf{Inst. Opt.} \\ \hline\Tstrut
Input MRI sequences            &   T1ce     &   T1,T1ce,T2,Flair    &     T1ce,T2       \\
Number of levels $N_{\text{level}}$            &   3     &   3    &     5       \\
Max. image dimension $N_{\text{max}}$         &   (64, 64, 40)     & (160, 160, 80)  &  (240, 240, 155)     \\
Min. image dimension $N_{\text{min}}$        &    (16, 16, 16)    &   (40, 40, 20)    & (80, 80, 80)       \\
Learning rate per level &    [1e-2, 5e-3, 2e-3]    &   [1e-4, 1e-4, 1e-4]    &   [1e-2, 5e-3, 5e-3, 3e-3, 3e-3]         \\
Max. iteration per level $N_{\text{iter}}$ &    [90, 90, 90]    &   -    &     [150, 100, 100, 100, 50]       \\
Distance measure             &    NGF($\epsilon=0.01$)    &   NCC-SP($w=7$)    & NCC($w=3$) \\
Max. number of grid points             &    -    &   $160\times160\times80$    & $64\times64\times64$ \\
Min. number of grid points             &    -    &   $40\times40\times20$    & $32\times32\times32$ \\
Weight of Inverse consistency $\lambda_{inv}$             &    -    &   [0.5, 0.5, 0.5]    & [1.0, 2.0, 4.0, 8.0, 10.0] \\
\end{tabular}
}
\end{table}

\subsubsection{Forward-Backward Consistency}
Given the deformation fields $\phi_{bf} = Id + \bm{u}_{bf}$ and $\phi_{fb} = Id + \bm{u}_{fb}$, where $Id$ is the identity transform and $\bm{u}$ is the corresponding displacement vector field, we calculate the forward-backward error $\delta_{bf}$ for $B$ to $F$ as:

\begin{equation}\label{eq:forward_backward}
	\delta_{bf}(x) = |\bm{u}_{bf}(x) + \bm{u}_{fb}(x+\bm{u}_{bf}(x))|_2.
\end{equation}

\noindent Based on the observation that regions without true correspondences would have higher forward-backward error in solutions, we create a mask $m_{bf}$ and mark $m_{bf}(x)=1$ whenever the forward-backward error $\delta_{bf}(x)$ is larger than the pre-defined threshold $\tau_{bf}(x)$. The pre-defined threshold is defined as:

\begin{equation}\label{eq:threshold}
	\tau_{bf} = \sum_{x\in\{x | F(x)>0\}} \frac{1}{N_f} \big(|\bm{u}_{bf}(x) + \bm{u}_{fb}(x+\bm{u}_{bf}(x))|_2\big) + \alpha,
\end{equation}

\noindent where $\alpha$ is set to 0.015. Then, we create a binary mask $\bm{m}_{bf}$ to mark voxels with absent correspondence as follows: 
\begin{equation}\label{eq:binary_mask}
    \bm{m}_{bf}(x)= 
\begin{cases}
    1,& \text{if } (\bm{A} \star \delta_{bf})(x) \geq \tau_{bf}\\
    0,              & \text{otherwise}
\end{cases}
\end{equation}

\noindent where $\bm{A}$ denotes an averaging filter of size $(2p+1)^3$ and $\star$ denotes a convolution operator with zero-padding $p$. 

\subsubsection{Objective Function}
The objective function $\mathcal{L}$ of DIRAC is defined as follows: 

\begin{equation}\label{eq:objective}
	\mathcal{L} = (1-\lambda_{reg})\mathcal{L}_{\text{s}} + \lambda_{reg}\mathcal{L}_{\text{r}} + \lambda_{inv}\mathcal{L}_{\text{inv}} + \lambda_{m}(|m_{bf}|+|m_{fb}|)
\end{equation}

\noindent where the masked dissimilarity measure $\mathcal{L}_{\text{s}}$ and the masked inverse consistency loss $\mathcal{L}_{inv}$ are defined as:
\begin{equation}\label{eq:dissim}
\mathcal{L}_{\text{s}}=\mathcal{L}_{\text{sim}}(B, F(\phi_{fb}), (1-m_{fb})) + \mathcal{L}_{\text{sim}}(F, B(\phi_{bf}), (1-m_{bf}))
\end{equation}
\noindent and
\begin{equation}\label{eq:inv}
\mathcal{L}_{\text{inv}}=\sum_{x\in\Omega} (\delta_{bf}(x)(1-m_{bf}(x)) + \delta_{fb}(x)(1-m_{fb}(x))).
\end{equation}

In this step, we use masked negative local cross-correlation (NCC) with similarity pyramid \cite{mok2020large} as the dissimilarity function. To encourage smooth solutions and penalize implausible solutions, we adopt a diffusion regularizer $\mathcal{L}_{\text{r}}=||\nabla \bm{u}_{bf}||^2_2+||\nabla \bm{u}_{fb}||^2_2$ during training. We set $\lambda_{reg}$ and $\lambda_{m}$ to 0.4 and 0.01, respectively. For more details, we recommend interested reader also refer to \cite{mok2022unsupervised}.

\begin{figure}[t]
	\centering
	\includegraphics[width=1.0\linewidth]{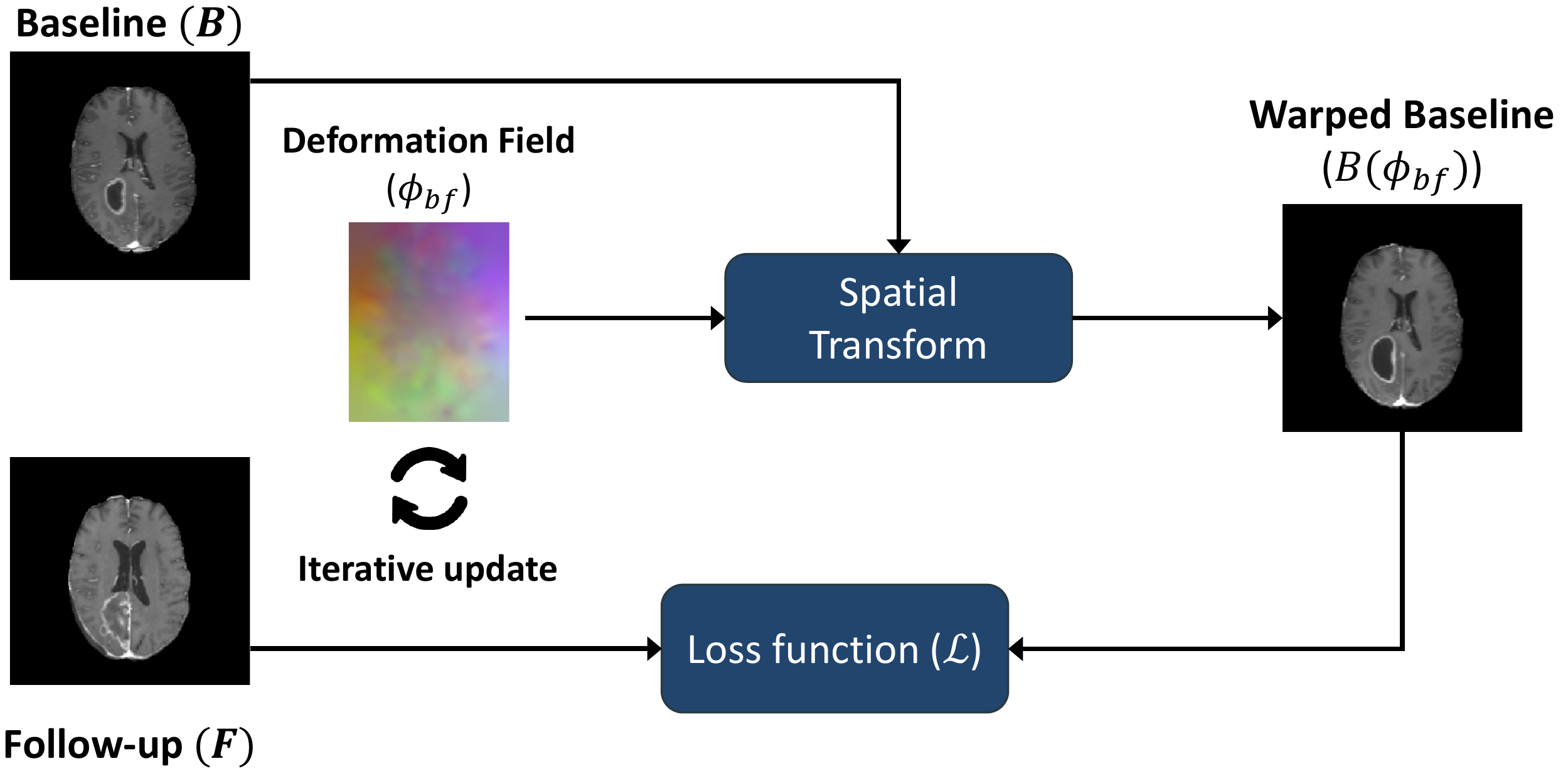}
	\caption{Overview of the non-rigid instance optimization (Step 3). Initially, the deformation field is initialized with the solution estimated from Step 2. The bidirectional deformation fields are jointly updated using Adam optimizer. For brevity, only baseline scan register to the follow-up scan is shown in the figure.} \label{fig:nonrigid}
\end{figure}

\begin{table*}[t]
	\centering	
	\caption{Results on the training set in the BraTS-Reg challenge. MAE$_\text{median}$ and MAE$_\text{mean}$ denote the average of median absolute error and mean absolute error of the transformed coordinates and the manually defined coordinates (in millimetre), respectively. Initial: Results before registration.}\label{tab:training}
	
	\renewcommand{\arraystretch}{0.95}
	\newcolumntype{C}[1]{>{\centering\arraybackslash}p{#1}}
	\newcolumntype{L}[1]{>{\arraybackslash}p{#1}}
	\newcommand*{\MyIndent}{\hspace*{0.35cm}}%
	\begin{tabular}{L{2.2cm} C{2.6cm} C{2.2cm} C{2.2cm}}
		\toprule
		\MyIndent Methods &  MAE$_\text{median}$ & MAE$_\text{mean}$ & Robustness\\
		\midrule
        \MyIndent Initial & $8.20 \pm 7.62$ & $8.65 \pm 7.31$  & - \\
        \midrule
        \MyIndent Ours & $2.98\pm 6.25$ & $4.64 \pm 11.06$ & $0.78 \pm 0.23$ \\
		\bottomrule
	\end{tabular}

\end{table*}

\subsection{Non-rigid Instance Optimization}
Due to insufficient amount of training data and discrepancy in distributions between training and test set, the learning-based method in step 2 may produce biased solutions, especially in cases with small initial misalignment. As such, we introduce a non-rigid instance optimization step to further improve the robustness and registration accuracy of solutions from the previous step. Figure \ref{fig:nonrigid} shows the overview of the non-rigid instance optimization. In the final step, the non-parametric deformation is controlled by the same objective function in \ref{eq:objective}, except we use NCC as the distance measure. The smoothness regularization coefficients $\lambda_{reg}$ for each level are set to [0.25, 0.3, 0.3, 0.35, 0.35], respectively. The displacement fields are discretized with trilinear interpolation defined on a uniform control point grid with a fixed number of points. We use an Adam optimizer together with multi-level continuation to avoid local minima. 

\subsection{Hyperparameters}
The hyperparameters of our 3-step approach are summarized in Table \ref{tab:parameters}.

\begin{table*}[t]
	\centering	
	\caption{Results on the validation set in the BraTS-Reg challenge. MAE and Robustness denote the average of median absolute error and mean absolute error of the transformed coordinates and the manually defined coordinates (in millimetre), respectively. Robustness measures the successful-rate of the registered landmarks. Affine, A-DIRAC and A-DIRAC-IO denote affine registration, DIRAC with affine pre-alignment and our proposed 3-step method, respectively. Initial: Results before registration. }\label{tab:validation}
	
\begin{tabular}{c|c|cccccc}
\multicolumn{1}{C{2.cm}|}{\multirow{2}{*}{Case}} & \multicolumn{1}{C{1.5cm}|}{Initial} & \multicolumn{2}{c}{Affine}                               & \multicolumn{2}{c}{A-DIRAC}                              & \multicolumn{2}{c}{A-DIRAC-IO}                           \\ \cline{2-8}
\multicolumn{1}{c|}{}                      & \multicolumn{1}{c|}{MAE}     & \multicolumn{1}{C{1.0cm}}{MAE\Tstrut} & \multicolumn{1}{c}{Robustness} & \multicolumn{1}{C{1.0cm}}{MAE} & \multicolumn{1}{c}{Robustness} & \multicolumn{1}{C{1.0cm}}{MAE} & \multicolumn{1}{c}{Robustness} \\ \hline\Tstrut
Case 141 & 13.50 & 4.26 & 1.00 & 1.94 & 1.00 & 1.62 & 1.00 \\
Case 142 & 14.00 & 6.12 & 0.88 & 3.07 & 1.00 & 1.88 & 1.00 \\
Case 143 & 16.00 & 8.98 & 1.00 & 2.63 & 1.00 & 1.14 & 1.00 \\
Case 144 & 15.00 & 9.52 & 0.88 & 3.10 & 1.00 & 2.56 & 1.00 \\
Case 145 & 17.00 & 5.14 & 1.00 & 2.09 & 1.00 & 1.13 & 1.00 \\
Case 146 & 17.00 & 5.74 & 1.00 & 1.84 & 1.00 & 1.94 & 1.00 \\
Case 147 & 1.50 & 2.03 & 0.45 & 2.17 & 0.60 & 1.64 & 0.55 \\
Case 148 & 3.50 & 2.90 & 0.75 & 1.71 & 0.90 & 1.43 & 0.90 \\
Case 149 & 9.00 & 2.22 & 1.00 & 1.94 & 1.00 & 1.56 & 1.00 \\
Case 150 & 4.00 & 3.65 & 0.53 & 2.87 & 0.63 & 1.27 & 0.74 \\
Case 151 & 3.00 & 2.13 & 0.5 & 1.39 & 0.75 & 1.18 & 0.85 \\
Case 152 & 5.00 & 2.11 & 0.95 & 1.45 & 0.84 & 1.42 & 0.95 \\
Case 153 & 2.00 & 2.04 & 0.33 & 1.44 & 0.67 & 1.80 & 0.67 \\
Case 154 & 2.00 & 2.61 & 0.25 & 2.02 & 0.4 & 1.98 & 0.55 \\
Case 155 & 2.00 & 3.09 & 0.21 & 2.43 & 0.37 & 1.70 & 0.53 \\
Case 156 & 7.00 & 2.84 & 1.00 & 2.29 & 1.00 & 1.45 & 1.00 \\
Case 157 & 10.00 & 4.90 & 0.90 & 2.67 & 1.00 & 1.66 & 1.00 \\
Case 158 & 4.50 & 3.48 & 0.40 & 1.39 & 0.80 & 1.13 & 1.00 \\
Case 159 & 6.00 & 7.28 & 0.36 & 2.25 & 1.00 & 2.28 & 1.00 \\
Case 160 & 4.00 & 2.55 & 0.7 & 2.29 & 0.80 & 1.94 & 0.90 \\
\hline\Tstrut
Mean & 7.80 & 4.18 & 0.70 & 2.15 & 0.84 & 1.64 & 0.88 \\
Std & 5.62 & 2.30 & 0.29 & 0.54 & 0.21 & 0.39 & 0.17 \\
Median & 5.50 & 3.28 & 0.81 & 2.13 & 0.95 & 1.63 & 1.00 \\
\end{tabular}

\end{table*}

\begin{figure}[t]
	\begin{center}
		\includegraphics[width=0.95\linewidth]{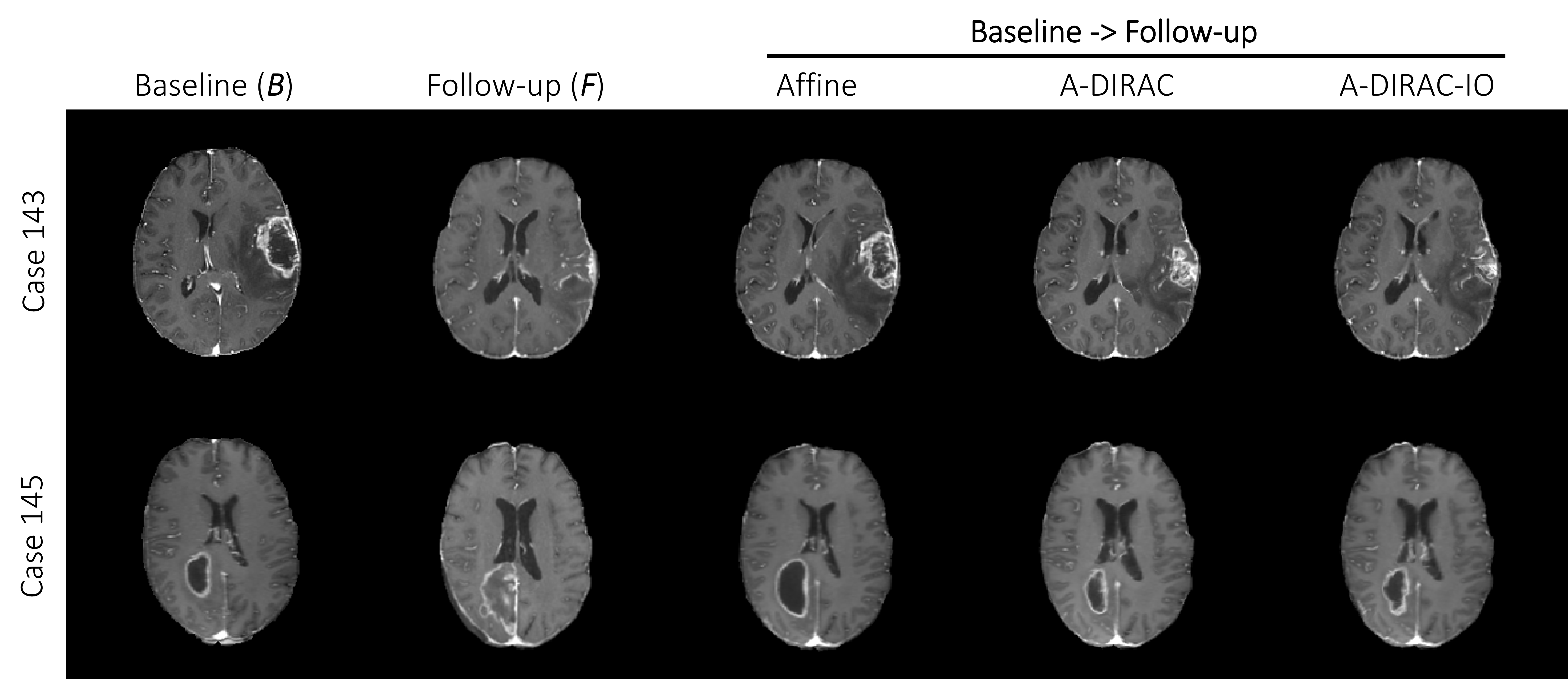}
	\end{center}
	
	\caption{Example axial T1ce MR slices of resulting warped images ($B$ to $F$) from affine, A-DIRAC and A-DIRAC-IO registration methods.}\label{fig:qualitative}

\end{figure}

\section{Experiments}
\subsubsection{Implementation} 
Our proposed method is implemented with PyTorch 1.8 \cite{paszke2017automatic} and trained with an Nvidia Titan RTX GPU and an Intel Core (i7-4790) CPU. We build DIRAC on top of the official implementation of cLapIRN available in \cite{offical_clapirn}. We adopt Adam optimizer \cite{kingma2014adam} with a fixed learning rate $1\mathrm{e}{-4}$ and train it from scratch with the training data from the BraTS-Reg challenge.



\subsubsection{Measurements}
We quantitatively evaluate our method based on the average of the median absolute error (MAE) and robustness of anatomical landmarks. Specifically, the MAE is defined as:

\begin{equation}\label{eq:mae}
	\text{MAE} = \text{Median}_{l \in L}(|x_l^B - \hat{x}_l^B|),
\end{equation}

\noindent where $x_l^B$ is the $l$-th estimated anatomical landmark in the baseline scan and $\hat{x}_l^B$ is the $l$-th goundtruth landmark in the baseline scan. The robustness is defined as the ratio of landmarks with improved MAE after registration to all landmarks, following the definition in \cite{baheti2021brain}. 

\subsubsection{Results}
For each case in the training and validation set of the BraTS-Reg challenge, we register the pre-operative image scan to the follow-up image scan and use the resulting deformation field to transform the manually defined landmarks in the follow-up scan. In total, there are 140 and 20 pairs of pre-operative and follow-up image scans in the training and validation set, respectively. We follow the evaluation pipeline of the BraTS-Reg challenge and report the average median absolute error MAE and robustness of the training and validation set in Tables \ref{tab:training} and \ref{tab:validation}. 

An example qualitative result is shown in Figure \ref{fig:qualitative}. The reduction of the registration error in the validation set in the pipeline is shown in Table \ref{tab:validation}. While the MRI scans are pre-registered to a common template, the average median error is reduced from 7.8 mm to 4.18 mm, indicating there exists a large linear misalignment between each case. Furthermore, the median error and robustness are consistently improved after each step, reaching to 1.64 mm average median error. Notably, our MAE is the lowest on the challenge's validation leaderboard. 



\section{Conclusion}
We proposed a 3-step registration method for pre-operative and follow-up brain tumor registration. The method was evaluated with the dataset provided by the BraTS-Reg challenge and ranked 1$^\text{st}$ place in the 2022 MICCAI BraTS-Reg challenge. By combining the pathological-aware deep learning-based method and instance optimization, we demonstrated the follow-up scan could be accurately registered to the pre-operative scan with an average median absolute error of 1.64mm. Compared to conventional methods, our method inherits the runtime advantage from deep learning-based approaches and does not require any manual interaction or supervision, demonstrating immense potential in the fully-automated patient-specific registration. We left the further analysis of our method and the comparison to existing methods for future work.

\bibliographystyle{splncs04}
\bibliography{myref}

\end{document}